\documentclass[aps,prb,twocolumn,showpacs,groupedaddress]{revtex4}

\usepackage{graphicx,bm}
\usepackage{amsmath,amssymb}
\usepackage{hyperref}
\usepackage{graphicx,amsfonts,amsbsy}
\usepackage{dsfont}

\begin{document}

\title{Rashba and intrinsic spin-orbit interactions in biased bilayer graphene}

\author{Ralph van Gelderen}
\email{R.vangelderen1@uu.nl}

\affiliation{Institute for Theoretical Physics, Utrecht
University, Leuvenlaan 4, 3584 CE Utrecht, The Netherlands}

\author{C. Morais Smith}

\affiliation{Institute for Theoretical Physics, Utrecht
University, Leuvenlaan 4, 3584 CE Utrecht, The Netherlands}

\date{\today}


\begin{abstract}
We investigate the effect that the intrinsic spin-orbit and the inter- and intra-layer
Rashba interactions have on the energy spectrum of either an unbiased or a biased
graphene bilayer. We find that under certain conditions, a Dirac cone is formed
out of a parabolic band and that it is possible to create a "Mexican hat"-like energy
dispersion in an unbiased bilayer. In addition, in the presence of only an
intralayer Rashba interaction, the $K$ $(K')$ point splits into four distinct ones,
contrarily to the case in single-layer graphene, where the splitting also takes place,
but the low-energy dispersion at these points remains identical.
\end{abstract}
\vskip2pc

\maketitle

\section{Introduction}

The influence of spin-orbit (SO) interactions on a single layer of graphene is well known. Kane and Mele \cite{KaMe05a,KaMe05b} were the first who showed that the intrinsic SO (ISO) interaction not only can open a gap, but it also gives rise to a quantum spin Hall phase, due to localized edge states. On the other hand, the extrinsic Rashba SO interaction acts in the opposite direction, and tends to close the gap. Later, it was found that the Rashba interaction has interesting effects on its own, leading to a splitting of the Dirac point into four identical points.\cite{ZaSa09} This splitting is missed in the low energy calculations used in Refs.~\onlinecite{KaMe05a,KaMe05b}.

The presence of a gap controlled by the ISO interaction seemed promising, but it turned out to be smaller than originally expected by Kane and Mele.\cite{Yao07,ZaSa07} The value of its coupling constant is still controversial, but is expected to be in the range $0.0011-0.05$ meV,\cite{Tri07,BoTr07} thus very small. The Rashba coupling instead, can be tuned to much higher values. For typical values of an external electric field ($50 \textrm{ V}/300 \textrm{ nm}$), the Rashba coupling is less than $1$ meV.\cite{MacDo06,Bra06} The effect of impurities can increase this value to $7$ meV.\cite{CaNe09b} However, recent experiments on epitaxial graphene grown on a Ni(111) substrate showed that the Rashba coupling can reach values up to $0.2$ eV.\cite{Yu08}

Shortly after the discovery of graphene, it was observed that bilayer graphene also exhibits remarkable phenomena. In bilayers, the low energy excitations are no longer Dirac fermions like in graphene, but massive chiral fermions.\cite{No06} In addition, bilayer graphene turns out to be a semiconductor, with a gap that can be tuned via a chemical doping\cite{Oh06} or by an external gate voltage.\cite{CaNe06,CaNe07,CaNe08,Oo07,McCa06}

\begin{figure}
\includegraphics[width=.45\textwidth]{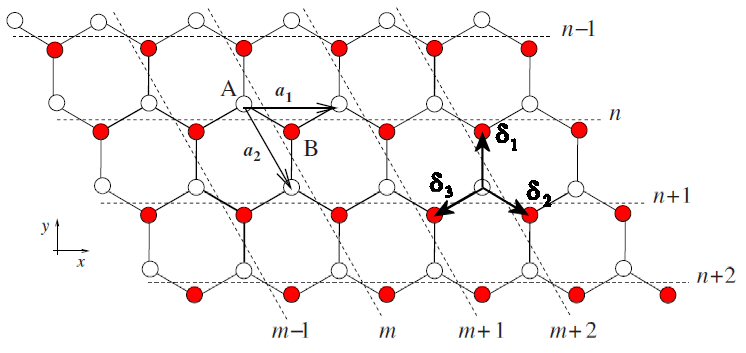}
\caption{Relabeling of the graphene lattice. The length of the lattice vectors shown equals $a$, while the lattice spacing is $a/\sqrt{3}$. Figure extracted and modified from Ref.~\onlinecite{CaNe09}.}
\label{relgraphlat}
\end{figure}

Although there are several studies for the effect of SO interactions in graphene accounting for different boundaries\cite{ZaSa09b,ZaBuSa08} (zigzag and armchair) and electron-electron interactions,\cite{ZaSa08} no investigations of the SO effects have so far been performed in bilayer graphene, to the best of our knowledge. In this paper, we incorporate SO interactions in a bilayer graphene system in the presence and absence of a bias voltage. We distinguish between the ISO interaction, which respects the lattice symmetry and the extrinsic Rashba interaction, which is only present if the lattice symmetry is broken. Here, we break the lattice symmetry by introducing an electric field. Depending on the orientation of this electric field, there can be both intralayer as well as interlayer Rashba interactions. Furthermore, we consider how the energy spectrum deforms if the layer is biased with an electrical voltage.

As a main result, we find that the intralayer Rashba coupling in an unbiased bilayer not only splits the $K$ ($K'$) points into four, in a different way than it does for monolayer graphene,\cite{ZaSa09} but also creates a Dirac cone out of a parabolic band. In addition, we show that a fully spin-polarized Mexican-hat band arises in the energy spectrum of an unbiased layer, purely due to SO interactions.

The Mexican-hat-like dispersion appears in a variety of physical systems. This kind of spectrum, with a line of degenerate low-energy points forming a ring, was first discussed by Brazovskii,\cite{Bra75} who showed that it leads to a "weak" crystallization transition. In cold atoms physics, a Mexican-hat-like dispersion appears and gives rise to topologically different ground states in SO Bose-Einstein condensates.\cite{Gal08} The Mexican hat is known in high energy physics as well, where, for example, the Higgs mechanism is expected to be responsible for the mass of the vector bosons. For bilayer graphene without SO interactions, the energy dispersion has the Mexican-hat shape if the layers are biased. In this case the dispersion was shown to potentialize electron-electron interactions, thus leading to a ferromagnetic instability.\cite{CaNe07b}

\begin{figure}
\includegraphics[width=.45\textwidth]{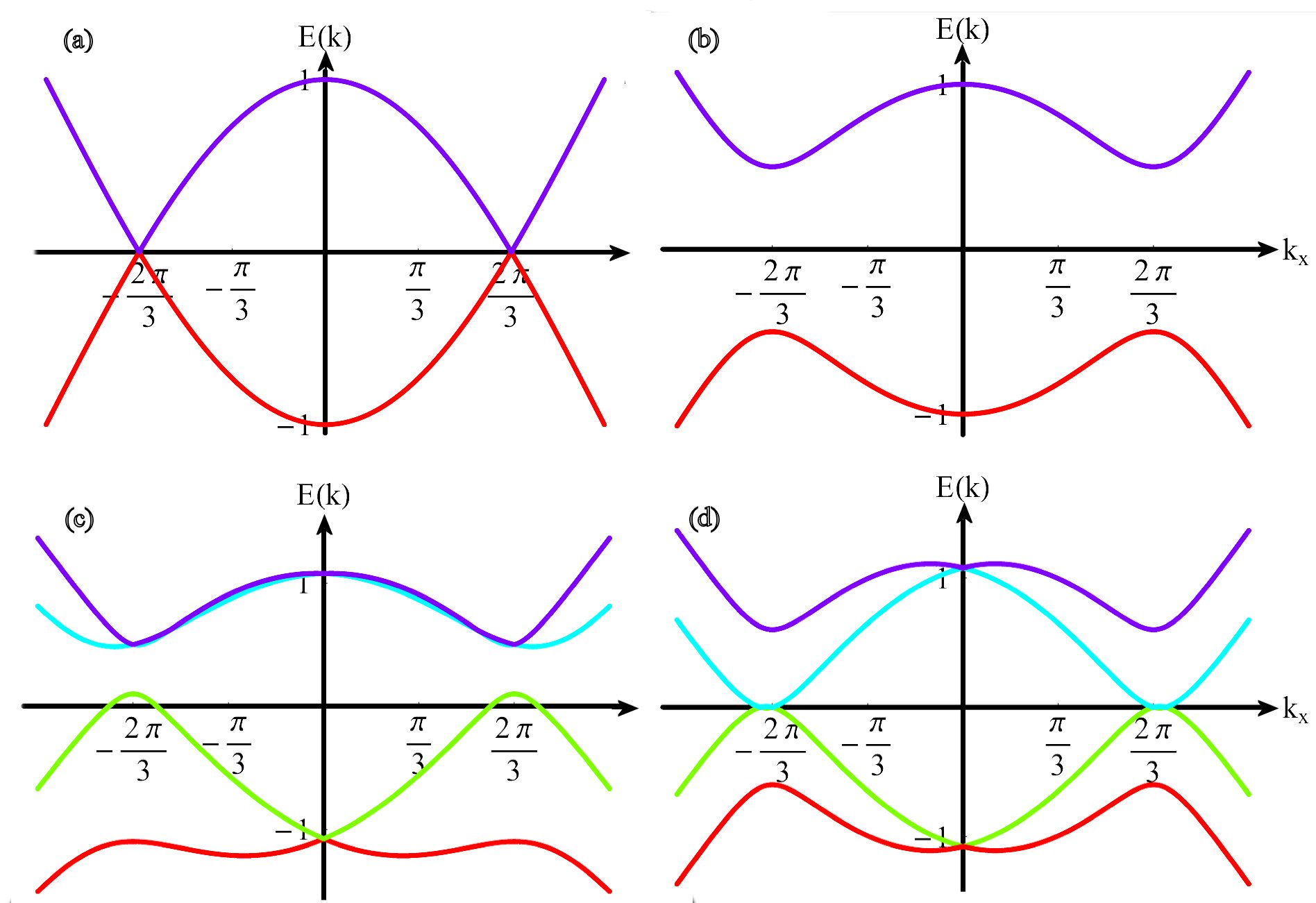}
\caption{(Color online) Behavior of a graphene sheet for different values of the SO parameters. (a) $\Delta_{SO}=0$, $t_R=0$. (b) $\Delta_{SO}=0.5$, $t_R=0$. (c) $\Delta_{SO}=0$, $t_R=0.2$. (d) $\Delta_{SO}=0.5$, $t_R=0.2$. In all figures $t=1$, $a=1$, $k_y=2 \pi \sqrt{3}/(3a)$.}
\label{monooverzicht}
\end{figure}

The outline of this paper is the following: To render the comparison with a monolayer sample easier, we recall some results for monolayer graphene in Sec.~\ref{monolayer}. Next, we set up a model for bilayer graphene in Sec.~\ref{bilayermodel}, after which we include SO interactions within the layers in Sec.~\ref{bilayerintraso}. We also include Rashba interactions between the layers. This is done in Sec.~\ref{bilayerintrainterso}. Finally, we draw our conclusions in Sec.~\ref{conclusions}.

\section{SO interactions in monolayer graphene}
\label{monolayer}

In graphene, the carbon atoms arrange themselves in a honeycomb lattice. Because there are two inequivalent positions for the carbon atoms, this honeycomb lattice can be seen as a triangular lattice with two atoms per unit cell, called $A$ and $B$, see Fig.~\ref{relgraphlat}. In the tight-binding approach, one assumes that the electrons are localized around the lattice sites and that they can hop from one lattice site to the next (nearest neighbor hopping). The noninteracting Hamiltonian is then given by \begin{align} \label{nonintham} H_0=t \sum_{ \substack{ i \in \Lambda_A,\sigma \\ j=1,2,3}} \left[ a_\sigma^\dag(\mathbf{R}_i)b_\sigma(\mathbf{R}_i+\boldsymbol{\delta}_j)+h.c \right], \end{align} where $\sigma$ is the spin index, $i$ runs over the $A$-sublattice sites and $j$ over the nearest neighbor vectors, which with a lattice orientation as in Fig.~\ref{relgraphlat}, are defined by
\begin{align}
\nonumber \boldsymbol{\delta}_1 &= \frac{a}{\sqrt{3}}\left(0,1 \right), \\ \nonumber \boldsymbol{\delta}_2 &= \frac{a}{\sqrt{3}}\left(\frac{\sqrt{3}}{2},-\frac{1}{2} \right), \\ \nonumber \boldsymbol{\delta}_3 &= \frac{a}{\sqrt{3}}\left(-\frac{\sqrt{3}}{2},-\frac{1}{2} \right).
\end{align}
The constant $t$ is the hopping parameter ($\approx2.8$ eV) and $a/\sqrt{3}\approx 0.142$ nm is the lattice spacing.

\begin{figure}
\includegraphics[width=.45\textwidth]{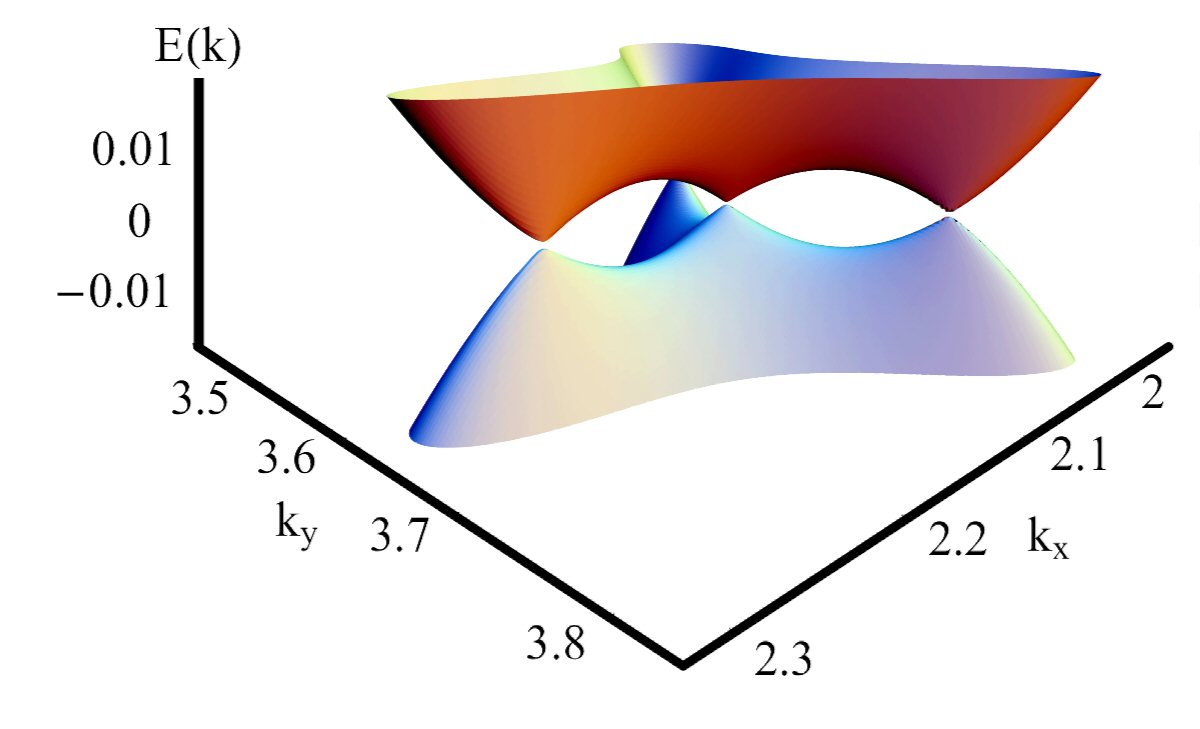}
\caption{(Color online) Splitting of the Dirac cones due to the Rashba interaction. For $\Delta_{SO}=0$ no gap opens. In this figure: $t=1$, $a=1$, $k_y=2 \pi \sqrt{3}/(3a)$, $\Delta_{SO}=0$, and $t_R=0.2$.}
\label{dconesplitting}
\end{figure}

After relabeling the lattice as in Fig.~\ref{relgraphlat}, one can bring the Hamiltonian \eqref{nonintham} into the form
\begin{align}
\nonumber H_0&=t \sum_{n,m,\sigma} \bigg[ a_\sigma^\dag(m,n) b_\sigma(m,n)+ a_\sigma^\dag(m,n) b_\sigma(m-1,n) \\
\nonumber &\phantom{=t \sum_{n,m}}+ a_\sigma^\dag(m,n) b_\sigma(m,n-1)+h.c. \bigg]. \end{align}
By performing a Fourier decomposition, the free Hamiltonian reads
\begin{align}
H_0 &= t\int d^2k \, \psi^\dag(k) M_{4 \times 4}^0  \psi(k),   \end{align}
where $\psi^\dag(k)=(a^\dag(k)_\uparrow,a^\dag(k)_\downarrow,b^\dag(k)_\uparrow,b^\dag(k)_\downarrow)$,
$$ M_{4 \times 4}^0 =\left( \begin{array}{cccc} 0 & 0 &  \gamma_\mathbf{k} & 0 \\ 0&0&0&  \gamma_\mathbf{k} \\ \gamma_\mathbf{k}^*  &0&0&0 \\ 0& \gamma_\mathbf{k}^* &0&0 \end{array} \right),$$
and $ |\gamma_\mathbf{k}|^2=3+2\cos(a k_x)+4\cos(a k_x/2)\cos(\sqrt{3}/2 a k_y).$
The eigenvalues of $H_0$ are the well known energy bands of graphene, \cite{CaNe09} $E_{\pm}=\pm t |\gamma_\mathbf{k}|$. Both bands are degenerate with respect to the spin degrees of freedom.

\begin{figure*}
\includegraphics[width=.95\textwidth]{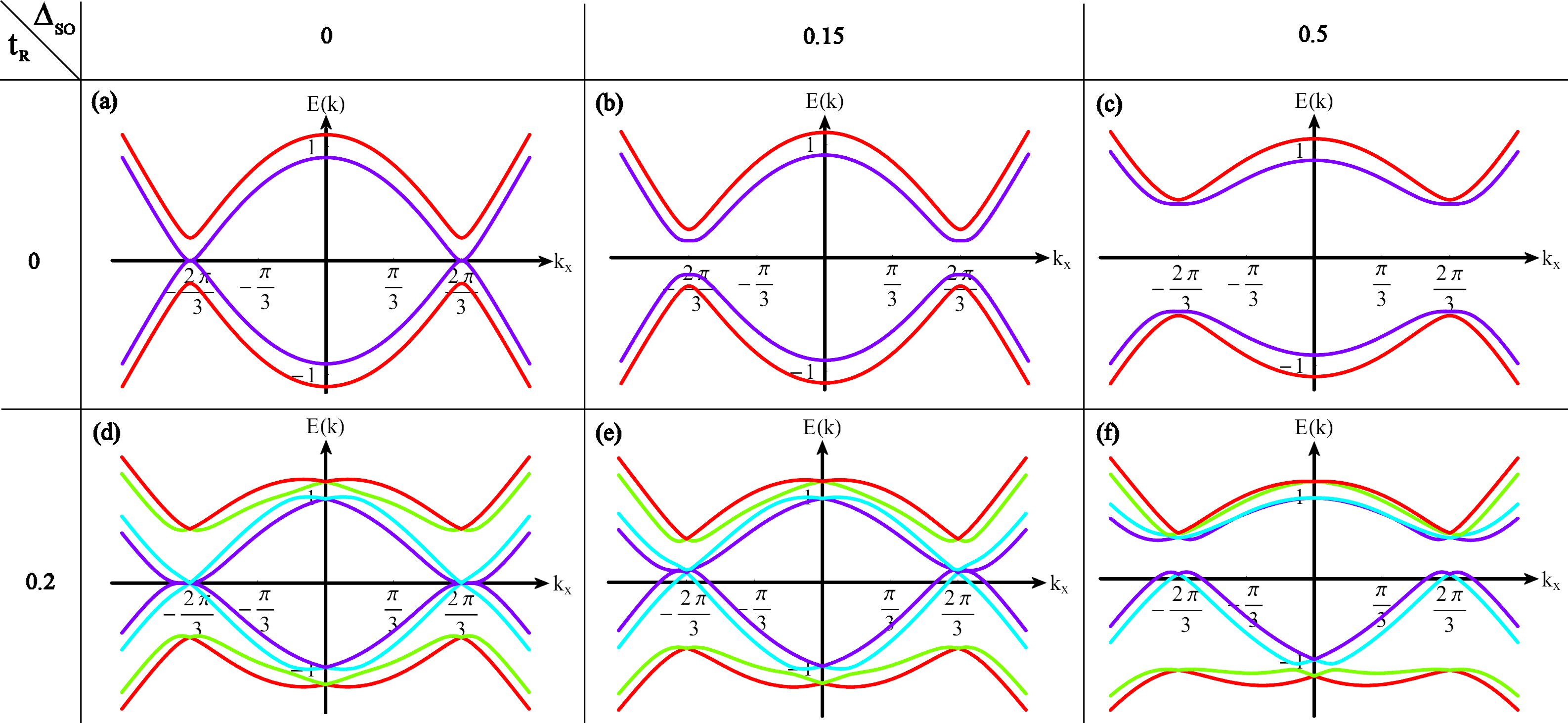}
\caption{ (Color online) Energy spectrum of bilayer graphene for different values of the SO interactions. In this figure the layer is unbiased, hence $V=0$. Other parameters have the values $t=1$, $t_\perp=0.2$, $a=1$, and $k_y=2 \pi/(\sqrt{3}a)$. }
\label{Vis0series}
\end{figure*}

If one wants to include SO interactions into the graphene system, one has to distinguish between the ISO interaction and the extrinsic Rashba term. The ISO interaction does respect all symmetries of the graphene lattice and has the form of a next-nearest neighbor (n.n.n.) hopping term,
\begin{align}
\nonumber H_{SO}=i t_{SO}\sum_{<<i,j>>}\nu_{ij} c^\dag_i s_z c_j.
\end{align}
In this expression, $t_{SO}$ is the n.n.n. hopping amplitude, $s_z$ is the $z$-component Pauli matrix describing the spin, and $c_j$ is either $a_j$ or $b_j$, depending wether the index $j$ labels an A or B-sublattice site, respectively. The factor $\nu_{ij}$ is $+1$ if the n.n.n. hopping is anti-clockwise and $-1$ if it is clockwise. Note that this term describes hopping within the same sublattice.

Using the relabeling of the lattice shown in Fig.~\ref{relgraphlat} and performing a Fourier decomposition, the ISO Hamiltonian can be rewritten as \begin{align}
\label{SOint} H_{SO}&= \Delta_{SO} \int d^2k \,  \psi^\dag(k) M_{4 \times 4}^{SO} \psi(k),  \end{align}
where
$$M_{4 \times 4}^{SO} =\left( \begin{array}{cccc} \eta & 0 & 0 & 0 \\ 0&-\eta&0& 0 \\ 0  &0&-\eta&0 \\ 0&0 &0&\eta  \end{array} \right), $$
$ \eta =[1/(3 \sqrt{3})] \left[ 2 \sin \left( a k_x \right)-4 \sin \left( a k_x/2 \right) \cos \left( \sqrt{3} a k_y/2 \right)\right]$ and $ \Delta_{SO}=3 \sqrt{3} \, t_{SO}$.

The extrinsic SO interaction is the Rashba term, which is only present if the lattice symmetry is broken. This can happen if the graphene sheet couples to a substrate or if an electric field is present. For a perpendicular electric field, $\mathbf{E}=E \, \hat{\mathbf{z}}$, the Rashba coupling has the form of a nearest neighbor hopping term and is given by \cite{ZaSa09,EnRaHa08,KaMe05b}
\begin{align}
\label{rashbaexpl} H_R &= i t_R \sum_{<i,j>} c^\dag_i \left( \mathbf{s} \times \hat{\mathbf{d}}_{ij} \right) \cdot \hat{\mathbf{z}} \, c_j+h.c.,
\end{align}
where the hopping amplitude $t_R$ can be tuned by changing the electric field strength, $\mathbf{s}$ is the vector of Pauli matrices and $\hat{\mathbf{d}}_{ij}$ is the unit vector that connects the $i$ and $j$ lattice sites. This term describes nearest neighbor hopping, but it only couples nearest neighbors with opposite spin. This is clearly seen if we rewrite this term in the same way that we rewrote the other terms,
\begin{align}
\label{Raint} H_{R}&= t_R \int d^2k \,  \psi^\dag(k) M_{4 \times 4}^R \psi(k),
\end{align}
where we have defined
\begin{align}
\nonumber M_{4 \times 4}^R &= \left( \begin{array}{cc}  0 & N_{2 \times 2} \\ N^\dag_{2 \times 2} & 0  \end{array} \right), \\
\nonumber N_{2 \times 2} &= \left( \begin{array}{cc} 0 & i\left[ \xi_1(k)+\xi_2(k)\right] \\  i\left[\xi_1(k)-\xi_2(k)\right] & 0 \end{array} \right), \\
\nonumber \xi_1(k) &=  e^{i \frac{1}{2} a k_x}  \left[ e^{-i \frac{\sqrt{3}}{2} a k_y}-\cos\left(\frac{1}{2} a k_x \right) \right], \\
\nonumber \xi_2(k) &= \sqrt{3} e^{i \frac{1}{2} a k_x} \sin \left( \frac{1}{2} a k_x \right).
\end{align}
The total Hamiltonian can be obtained by collecting Eqs. \eqref{nonintham}, \eqref{SOint}, and \eqref{Raint},
\begin{align}
\nonumber H=\int d^2k \, \psi^\dag(k) \left( t\, M_{4 \times 4}^0+\Delta_{SO} \, M_{4 \times 4}^{SO}+t_R \,M_{4\times 4}^R \right) \psi(k).
\end{align}

In Fig.~\ref{monooverzicht} we show the behavior of the energy dispersion for a graphene sheet for different values of the SO parameters, $\Delta_{SO}$ and $t_R$, in units of $t$. Without SO interactions, we find the well known graphene spectrum with Dirac cones centered at the $K$ and $K'$ points in the reciprocal space,\cite{CaNe09} see Fig.~\ref{monooverzicht}(a). The ISO interaction opens a gap, but respects the spin degeneracy of the energy bands (Fig.~\ref{monooverzicht}(b)). The Rashba term does not open a gap on its own, but it does lift the spin degeneracy, except at the $k_x=0$ point, as it can be seen in Fig.~\ref{monooverzicht}(c). By zooming in on the region around the $K$ ($K'$) point, we see that the Rashba term splits the Dirac cones into four, as noted in Ref.~\onlinecite{ZaSa09}. This behavior is depicted in Fig.~\ref{dconesplitting}. Note that this effect is missed in the approximation made by Kane and Mele,\cite{KaMe05b} which is effectively a zeroth order approximation of the Rashba term. If one takes linear terms in $k$ into account, this effect is already present. Here, however, we keep the full expression for the spectrum, without resorting to approximations. The combined effect of the ISO interaction and the Rashba term breaks the particle hole symmetry (Fig.~\ref{monooverzicht}(d)). If the Rashba term is small ($t_R<\Delta_{SO}$), the gap is finite, \cite{KaMe05b} otherwise the gap closes (not shown).

\begin{figure}
\includegraphics[width=.4\textwidth]{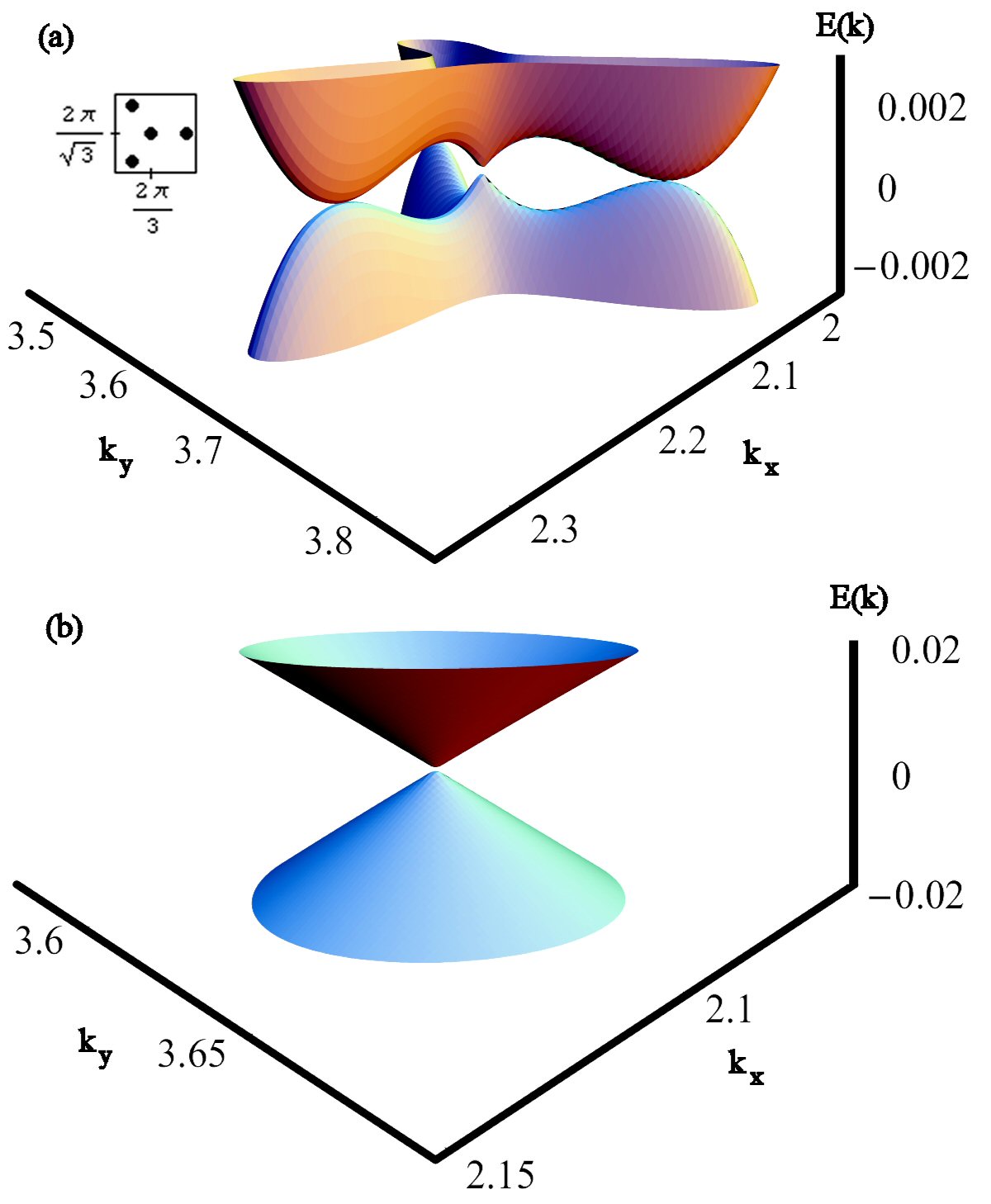}
\caption{(Color online) Due to the Rashba coupling, the spin degeneracy of the low laying energy bands is lifted. (a) One band exhibits four touching points between the valence and the conduction bands. Inset: points in $k$-space where the valence and the conduction bands touch. (b) The other band becomes a Dirac cone. In these figures, the layer is unbiased and there is no ISO interaction. Other parameters have the values $t=1$, $t_\perp=0.2$, $a=1$, and $k_y=2 \pi/(\sqrt{3}a)$.}
\label{dsplittingbilayer}
\end{figure}

\section{Bilayer graphene model}
\label{bilayermodel}

Before including the SO interactions into bilayer graphene, let us first consider the non-interacting Hamiltonian. The form of the Hamiltonian of a multilayer graphenesystem depends on the stacking of the layers.\cite{CaNe09} For bilayers, however, there are only two possibilities. The lattice sites can lay exactly on top of each other, or they can be arranged in a Bernal stacking, \footnote{Note that for bilayers there is no difference between Bernal stacking and Rhombohedral stacking, since they differ in the orientation of a possible third layer.} in which the $A$ sites of the upper layer ($A_1$) lay on top of the $B$ sites of the lower one ($B_2$), while the other sites ($B_1$ and $A_2$) lay opposite to a honeycomb center. We will assume the Bernal stacking here, because it is the most common one.

\begin{figure}[t]
\includegraphics[width=.4\textwidth]{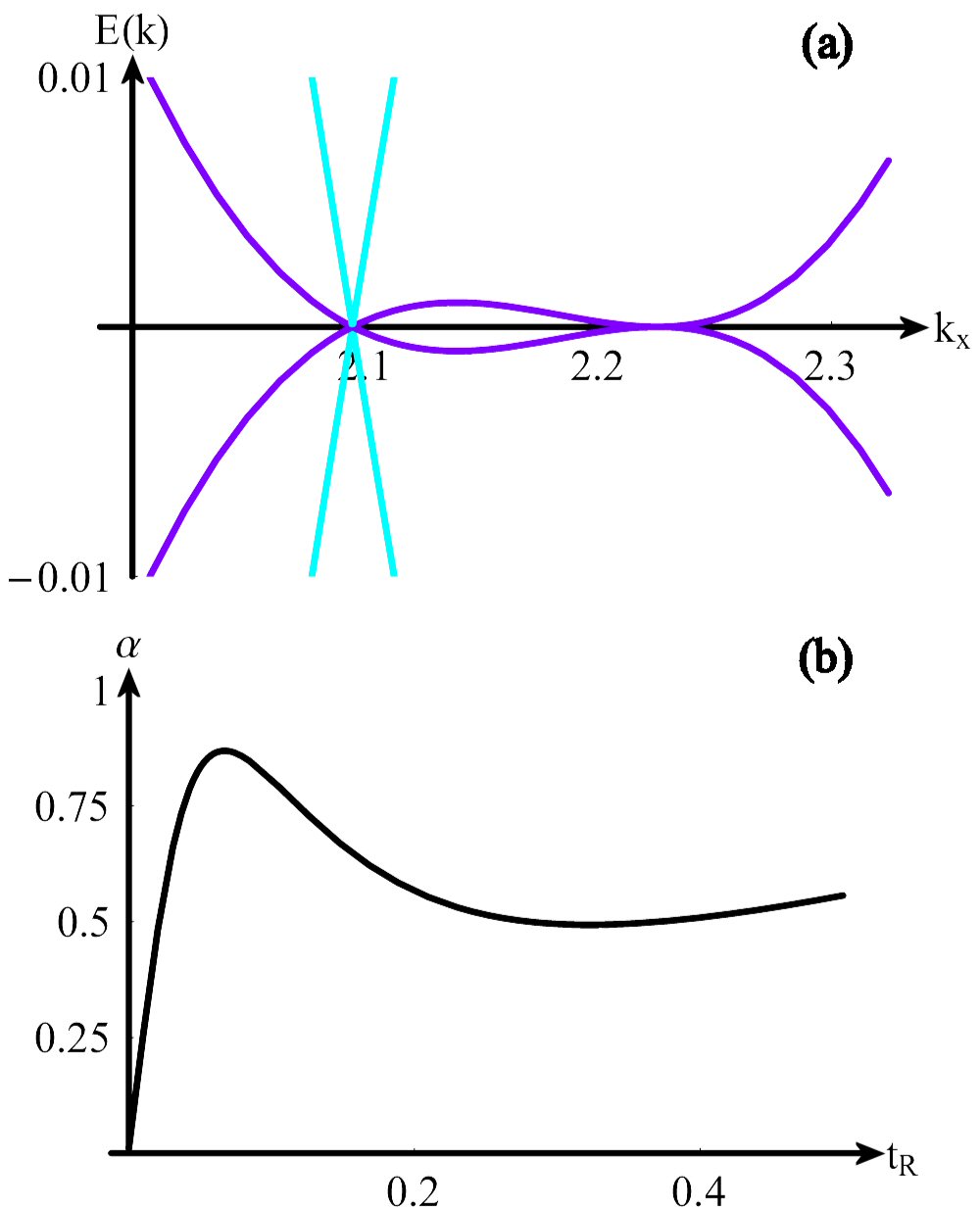}
\caption{(Color online) (a) Zoom in on the $K$ point of Fig.~\ref{Vis0series}(d) and an intersection along the line $k_y=2 \pi/(\sqrt{3}a)$ of Fig.~\ref{dsplittingbilayer}. (b) Slope ($\alpha$) of the Dirac cone as a function of the Rashba coupling $t_R$. The scale of the $\alpha$-axis depends on all parameters of the theory. Here: $t=1$, $t_\perp=0.2$, $a=1$, $V=0$, $\Delta_{SO}=0$, and $k_y=2 \pi/(\sqrt{3}a)$.}
\label{Vis0serieszi}
\end{figure}

\begin{figure*}
\includegraphics[width=.95\textwidth]{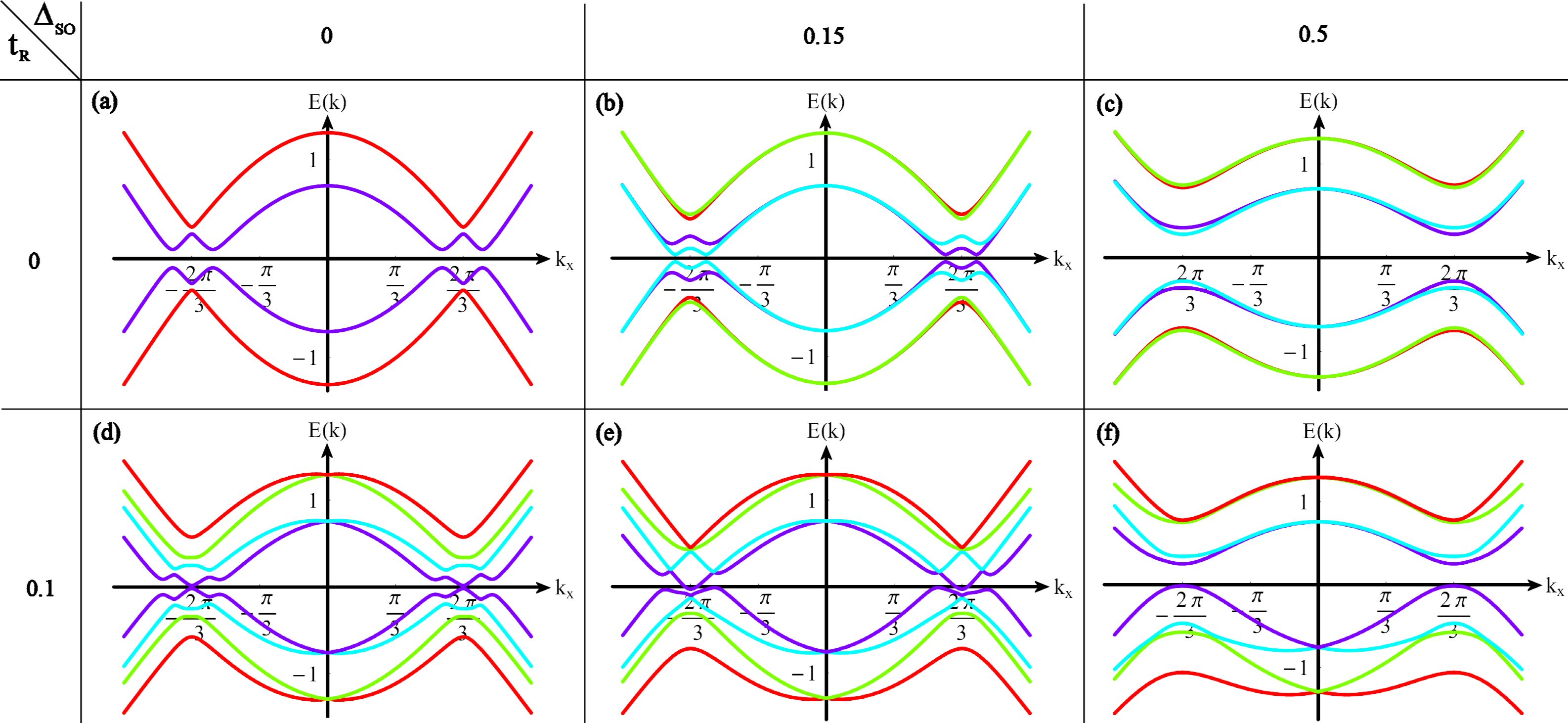}
\caption{(Color online) Energy spectrum of bilayer graphene for different values of the SO interactions. In this figure, the layer is biased, $V=0.25$. Other parameters have the values $t=1$, $t_\perp=0.2$, $a=1$, and $k_y=2 \pi/(\sqrt{3}a)$.}
\label{Vis025series}
\end{figure*}

The consequence of a Bernal stacking is that, in a first approximation, the only interlayer hopping is between $A_1$ and $B_2$ sites. It is straightforward to generalize the noninteracting monolayer Hamiltonian to a bilayer one. First, we define $$\Psi(k)^\dag=\left( a^\dag_{\uparrow,1} \, , \,a^\dag_{\downarrow,1} \, , \,b^\dag_{\uparrow,1} \, , \,b^\dag_{\downarrow,1} \, , \,a^\dag_{\uparrow,2} \, , \,a^\dag_{\downarrow,2} \, , \,b^\dag_{\uparrow,2} \, , \,b^\dag_{\downarrow,2} \right),$$ where the layers $1$ and $2$ are represented by the corresponding index.
We introduce an interlayer hopping parameter, $t_\perp \approx (0.1-0.2) \, t$, and bias the bilayer system with a gate voltage $V$. This gate voltage can be tuned externally and is such that the lower layer has an electric potential $-V$, while the upper layer has $V$. With this new parameters, the non-interacting Hamiltonian is given by
\begin{align}
H_0^{bl}&=\int d^2k \, \Psi^\dag(k) M^0_{8 \times 8} \Psi(k), \\
\nonumber M^0_{8 \times 8}&=\left( \begin{array}{cc} V \mathds{1}_{4\times4}+t \, M^0_{4 \times 4} & A \\ A^\dag & -V \mathds{1}_{4\times4}+t \,M^0_{4 \times 4} \end{array} \right), \\
\nonumber A&=\left( \begin{array}{cccc} 0&0&t_\perp &0 \\ 0&0&0&t_\perp \\0&0&0&0 \\0&0&0&0 \end{array} \right).
\end{align}

If we consider $V=0$ for the moment, we see in Fig.~\ref{Vis0series}(a) that the spectrum is different from the one for monolayer graphene. The dispersion at the $K$ ($K'$) points, where the valence and the conduction bands touch, is no longer linear, but parabolic. This means that in a low energy approximation the quasi particles become massive ($m \approx 0.054 m_e$).\cite{CaFa06} However, these particles are chiral with respect to sublattice pseudospin and are therefore massive chiral fermions,\cite{No06} which are new type of quasi particles, characteristic for bilayer graphene.

\begin{figure*}[t]
\includegraphics[width=.95\textwidth]{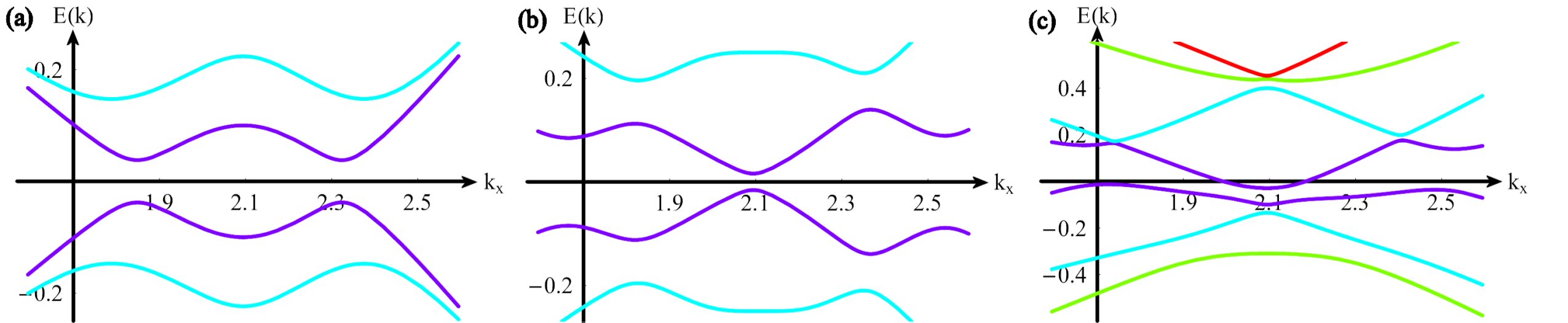}
\caption{(Color online) (a) Zoom in on the $K$ point of Fig.~\ref{Vis025series}(b). (b) Zoom in on the $K$ point of Fig.~\ref{Vis025series}(d). (c) Zoom in on the $K$ point of Fig.~\ref{Vis025series}(e). Recall that $V=0.25$, $t=1$, $t_\perp=0.2$, $a=1$, and $k_y=2 \pi/(\sqrt{3}a)$.}
\label{Vis025serieszi}
\end{figure*}

If there is a nonzero voltage difference, a gap will open in the energy spectrum. In fact, not only a gap opens, but the parabolic bands are deformed into Mexican hats (see Ref.~\onlinecite{CaNe06} and the discussion in Sec. \ref{bilayvolt}). However, the energy bands remain spin degenerate, because a nonzero voltage on its own cannot lift the spin degeneracy; as we will discuss in the next section, interactions are required to reach this aim.

\section{Intralayer SO interactions in bilayer graphene}
\label{bilayerintraso}

\subsection{No bias voltage}

Now that we have understood the single particle spectrum for the bilayer graphene system, let us add SO interactions. We already saw that we have to distinguish between the ISO and the Rashba interactions, but for the bilayer system there is another subdivision, namely into intralayer and interlayer interactions.
In this section, we will analyze the effects that the \textbf{intralayer} ISO and Rashba couplings have on the energy spectrum of the bilayer system. The effect of these interactions in the presence of a bias voltage are subsequently discussed.

The ISO interaction respects the symmetries of a single graphene sheet. Since a graphene bilayer has a smaller symmetry group than a single layer, we expect this interaction to be present in the planes of the bilayer system as well. The ISO interaction Hamiltonian is then given by
\begin{align}
H_{SO} &= \Delta_{SO} \int d^2k \, \Psi^\dag(k) M_{8 \times 8}^{SO} \Psi(k), \\
M_{8 \times 8}^{SO}&= \left( \begin{array}{cc} M_{4 \times 4}^{SO} & 0 \\ 0 & M_{4 \times 4}^{SO} \end{array} \right).
\end{align}

Regarding the Rashba term, we expect that a perpendicular electric field gives rise to intralayer interactions in the same way that it did for a single sheet of graphene. Effectively, we have two copies of the monolayer Rashba interaction,
\begin{align}
H_{R} &= t_R \int d^2k \, \Psi^\dag(k) M_{8 \times 8}^{R} \Psi(k), \\
M_{8 \times 8}^{R}&= \left( \begin{array}{cc} M_{4 \times 4}^{R} & 0 \\ 0 & M_{4 \times 4}^{R} \end{array} \right).
\end{align}

In Fig.~\ref{Vis0series}, the energy spectrum of an unbiased bilayer graphene is shown for different values of $\Delta_{SO}$ and $t_R$. For zero SO interactions, we observe the well known parabolic bands, which are spin degenerate (see Fig.~\ref{Vis0series}(a)). These degeneracies cannot be lifted by the ISO interaction on its own, which simply opens a gap in the spectrum (Fig.~\ref{Vis0series}(b) and (c)). This is the same behavior as for monolayer graphene. Things become interesting when we consider the case of zero ISO coupling and a nonzero Rashba interaction (Fig.~\ref{Vis0series}(d)). The spin degeneracy of the bands is then lifted, but in a very particular way. In Fig.~\ref{dsplittingbilayer}(a), we see that the $K$ point splits into four points again, as for monolayer graphene. One of them remains at the former $K$ point position and the others form a triangle around it, see the inset in Fig.~\ref{dsplittingbilayer}(a). We will refer to these four points as the split $K$ point. However, here very special features appear: Besides the two energy bands (conduction and valence band) that touch at four points, there are two more bands that touch and form a Dirac cone, as it can be seen in Fig.~\ref{dsplittingbilayer}(b). The center of this Dirac cone is exactly at the point in $k$ space where, without Rashba interaction, the $K$ point was located. This is also the location of the central of the four points that form the split $K$ point. This central point is different from the other three. If we analyze Fig.~\ref{Vis0serieszi}(a), which shows a zoom-in

of Fig.~\ref{Vis0series}(d) and a cross-section cut of Figs.~\ref{dsplittingbilayer}(a) and (b), we see that the central point, located at $k_x=2 \pi/(3a)$, $k_y=2 \pi/(\sqrt{3} a)$ has, in addition to the Dirac cone, a linear crossing at very low energy, whereas the off-center points have only a higher order crossing.
This is different from the case of monolayer graphene, where the $K$ point splits into four equivalent points. The most striking feature of the intralayer Rashba coupling is the formation of a Dirac cone out of a parabolic band. If we perform a low energy approximation and use $\mathbf{k}=\mathbf{K}+\mathbf{q}$, this Dirac cone has the dispersion relation $E(\mathbf{q})=\alpha |\mathbf{q}|$. The slope $\alpha$, which corresponds to the velocity of the low energy excitations, depends on the parameters of the theory and cannot be determined analytically. We have plotted the slope of this cone as a function of $t_R$ for certain parameter values in Fig.~\ref{Vis0serieszi}(b).

If we set both $t_R$ and $\Delta_{SO}$ unequal to zero (Figs.~\ref{Vis0series}(e) and (f)), we see that depending on their relative values, a gap can open. For small ISO interactions, the gap stays closed, but the bands are heavily deformed in an asymmetric manner.Moreover, the particle-hole symmetry is lost. The split $K$ point becomes so deformed that we cannot identify the four points any longer. If $\Delta_{SO}$ becomes large enough a gap opens, but the asymmetry remains.

\subsection{Effect of a bias voltage}
\label{bilayvolt}

\begin{figure*}
\includegraphics[width=1\textwidth]{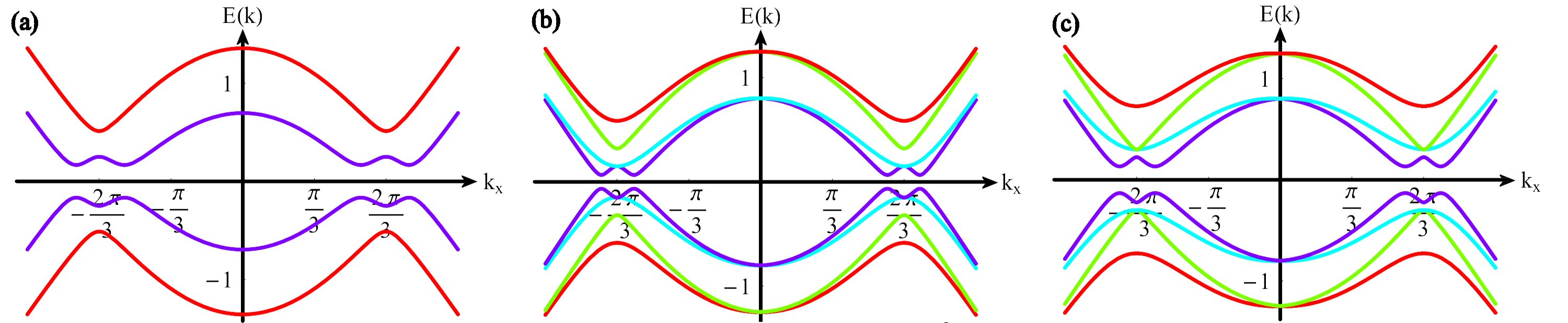}
\caption{(Color online) (a) Energy spectrum of a biased system, $V=0.25$, with zero intralayer SO interactions and $t_R^\perp=0.4$. Compare with Fig.~\ref{Vis025series}(a). (b) Energy spectrum of an unbiased bilayer system, with zero intralayer Rashba coupling, but with $\Delta_{SO}=0.15$ and $t_R^\perp=0.4$. Compare with Fig.~\ref{Vis0series}(b). (c) Energy spectrum of an unbiased bilayer system, with zero intralayer Rashba coupling, but with $\Delta_{SO}=0.3$ and $t_R^\perp=0.4$. Other parameters are $t=1$, $t_\perp=0.2$, $a=1$, and $k_y=2 \pi/(\sqrt{3}a)$.}
\label{trinter}
\end{figure*}

If we add a bias voltage to a bilayer graphene system without SO interactions, the system becomes a semiconductor with a tunable gap. We will see in the following that SO interactions can heavily deform the energy dispersion.

As found earlier,\cite{CaNe07, CaNe09} we observe that for $\Delta_{SO}=t_R=0$ the effect of the bias is to open a gap and to introduce a Mexican-hat-like shape in the lowest energy band around the $K$ and $K'$ points. The spectrum remains particle-hole symmetric and the bands remain spin degenerate, as is shown in Fig.~\ref{Vis025series}(a). By keeping $t_R=0$ and introducing a finite $\Delta_{SO}$, the spin degeneracy is lifted, but only in the region around the $K$ ($K'$) points, as can be seen in Fig.~\ref{Vis025serieszi}(a). When $\Delta_{SO}$ is not too large, the Mexican hat feature remains, but it disappears as $\Delta_{SO}$ is increased further (see Figs.~\ref{Vis025series}(b) and (c)). The gap at the $K$ point (this is not the actual gap if $V>\Delta_{SO}$, because of the Mexican hat structure), equals $2|V-\Delta_{SO}|$; Therefore, during the transition from the Mexican hat to a parabolic band the gap closes at $V=\Delta_{SO}$, after which it opens again. Note that this behavior is similar to that described in Ref.~\onlinecite{KaMe05a}, where both the ISO coupling $\Delta_{SO}$ and a staggered sublattice potential $\lambda_v$ can open a gap in monolayer graphene. If $\Delta_{SO}$ exceeds $\lambda_v$, a transition occurs between a normal insulator phase and a quantum spin Hall phase. Nonetheless, we have to be careful with this comparison, since a staggered sublattice potential in monolayer graphene is fundamentally different from a bias potential in a bilayer. For example, in monolayer graphene we do not observe a Mexican hat structure and the edge states are responsible for the phase transition. We do not take those into account here. In addition, although the ISO interaction alone cannot lift the spin degeneracy, it does so in the presence of a staggered sublattice potential in monolayer graphene. The lifting occurs for all values of $k$, except for $k=0$, contrarily to the biased bilayer, where it is only significant around the $K$ ($K'$) points. Moreover, the effect is stronger in monolayer graphene than it is in bilayer graphene.

As long as the Rashba coupling is zero, an analytical solution for the energy bands can be found. This solution is given by \begin{align} \nonumber E(k)^{\pm, \pm, \pm}&=\pm \frac{1}{\sqrt{2}} \bigg[ t_\perp^2+2 V^2+2 t^2 |\gamma|^2+2 \Delta_{SO}^2 \eta^2 \\ \nonumber &\phantom{= \pm} \pm \bigg(t_\perp^4+4 t^2 t_\perp^2 |\gamma|^2+16 t^2 V^2 |\gamma|^2 \\ \label{enbands} &\phantom{=\pm}+16 \Delta_{SO}^2 V^2 \eta^2 \pm 8 t_\perp^2 \Delta_{SO} V \eta \bigg)^{\frac{1}{2}} \bigg]^{\frac{1}{2}}. \end{align} It is clear that, if either $V$, $\Delta_{SO}$, or $t_\perp$ is zero the bands become degenerate, since in this case the last term

with the $\pm$ sign vanishes. Another interesting feature that we see only in a particular situation, where $t_\perp,V,\Delta_{SO} \neq 0$ and $t_R=0$, is a band crossing at $k_x=0$ (see Figs.~\ref{Vis025series}(b) and (c) and notice the inverted colors at the $K$ and $K'$ points), signalling that the $k_x \to -k_x$ symmetry is lost. This band crossing can be seen analytically from Eq.~\eqref{enbands}. Note that $|\gamma_k|^2$ is symmetric, while $\eta$ is antisymmetric under $k_x \to -k_x$. We conclude that, due to the linear term in $\eta$, the energy bands satisfies $E(k)^{\pm,\pm,\pm}=E(-k)^{\pm,\pm,\mp}$. Therefore, the individual bands are no longer symmetric under this transformation.

Now, let us investigate the behavior of the system at finite $t_R$, for $\Delta_{SO}=0$. First of all, the spin degeneracy is lifted, except at $k_x=0$. Second, the Mexican-hat feature evolves into something that looks more like an asymmetric farmers hat, see Fig.~\ref{Vis025series}(d) and Fig.~\ref{Vis025serieszi}(b). Note that if $V$ and/or $t_R$ are increased, the asymmetry becomes more accentuated and the spectrum does not even look like a farmers hat anymore.
We must emphasize that here we do not use any approximation for the energy spectrum, but we keep the full expression.
This feature cannot be captured in a zeroth order approximation for the Rashba term, as used by Kane and Mele \cite{KaMe05a,KaMe05b}, because at this order of the approximation the spectrum is symmetric around the $K$ ($K'$) point.

If both $\Delta_{SO}$ and $t_R$ are finite, the spectrum becomes very complicated (Figs.~\ref{Vis025series}(e) and (f)). As a general trend, $\Delta_{SO}$ washes out the Mexican hat feature and at first, increases the difference between spin up and spin down bands around the $K$ ($K'$) points, although the spin degeneracy had been already lifted for all values, except at $k_x=0$, by the finite Rashba coupling. In addition, particle-hole symmetry is lost. There is no longer a band crossing, but the gap closes and opens again upon increasing $\Delta_{SO}$. Depending on the parameters, a situation can occur where the bands do not touch, but they have common energies, thus there is no gap in the system (see Fig.~\ref{Vis025serieszi}(c)).

\section{Intra- and inter-layer SO interactions in bilayer graphene}
\label{bilayerintrainterso}

In the previous section we accounted for tunneling between the two layers, but considered only intralayer SO interactions. Now, we investigate the effect of SO interactions between lattice sites in different layers. Since the ISO interaction depends on the symmetry of the graphene plane, it is not obvious if there is any interplane ISO interaction at all. Hence, we focus on the Rashba term. This term only exists in graphene monolayers if the $z \to -z$ mirror symmetry is broken, for example by a perpendicular electric field. Furthermore, the Rashba coefficient can be tuned by varying this electric field. Therefore, we consider here a bilayer system in the presence of a tilted electric field. The inplane component of the electric field ($\mathbf{E}_\parallel$) gives rise to an interlayer Rashba coupling that is a generalization of Eq.~\eqref{rashbaexpl},
\begin{align}
\label{rashpar} H_R^\perp &=-i t_R^\perp \sum_i a^\dag_{i,1} \left( \mathbf{s} \times \hat{\mathbf{z}} \right) \cdot \hat{\mathbf{E}}_\parallel \, b_{i,2} + h.c.,\\
\nonumber \hat{\mathbf{E}}_\parallel &= \left( \cos \phi, \sin \phi, 0 \right)^T,
\end{align}
where we have chosen to absorb the magnitude of the electric field already in the constant $t_R^\perp$.
The orientation of $\mathbf{E}_\parallel$ is determined by $\phi$, but the results will be independent of $\phi$ and therefore we choose $\phi=0$ arbitrarily. The unit vector connecting the two lattice sites $A_{i,1}$ and $B_{i,2}$ is given by $-\hat{\mathbf{z}}$ and this explains the minus sign in comparison with Eq.~\ref{rashbaexpl}. In $k$-space, Eq.~\eqref{rashpar} becomes
$$ H_R^\perp= t_R^\perp \int d^2k \, \Psi^\dag(k) \, M_{8 \times 8}^{R, \perp} \, \Psi(k), $$
where the matrix $M_{8 \times 8}^{R, \perp}$ is given by
$$ M_{8 \times 8}^{R, \perp} = \left( \begin{array}{cc} 0 & C \\ C^\dag & 0 \end{array} \right), $$ with
$$ C= \left( \begin{array}{cccc} 0 & 0 & 0 & - e^{-i \phi} \\ 0 & 0 &  e^{i \phi}  & 0 \\ 0&0&0&0 \\ 0&0&0&0 \end{array} \right). $$
The effect of this interlayer Rashba interaction depends heavily on the other parameters in the theory. Without any intralayer SO interactions, the result of a nonzero $t_R^\perp$ is effectively a modification of the interlayer hopping parameter, \begin{align} \label{tperptrans} t_\perp \to t_\perp \sqrt{1+(t_R^\perp)^2/t_\perp^2}.\end{align}The effect is a slight deformation of the energy bands. Only if $t_{R}^\perp \gtrsim 0.3$ the shift becomes significant. For a biased system, $t_R^\perp$ will flatten the Mexican hat, as it can be seen in Fig.~\ref{trinter}(a).

The effect of interlayer Rashba coupling is most visible in bilayer systems with zero bias, but with intralayer ISO interactions. Although the energy spectrum can be solved analytically, the equations become too complicated to handle. However, it is clear that the effect is more than a shift of the interlayer hopping parameter. The spin degeneracy of the energy bands is lifted and we see a Mexican hat feature appear in the low laying energy band (see Fig.~\ref{trinter}(b)). These low energy bands are shifted towards the Fermi level and as a consequence the gap between the valence and the conduction bands becomes smaller. If $\Delta_{SO}$ is small, the spin degeneracy of the lowest energy band stays intact at the $K$ point, but is lifted around it. However, if $\Delta_{SO}$ increases, this degeneracy is lifted and eventually shifted to a degeneracy between the two conduction bands that lay in the middle, as it can be seen in Fig.~\ref{trinter}(c). In comparison with the bilayer without intralayer interactions, the effect of nonzero $t_R^\perp$ manifests itself also for small values of this parameter.
In a biased system with intralayer ISO interactions, the effect of the interlayer Rashba coupling is visible, but its influence becomes less important as the bias becomes larger. The effect of a nonzero $t_R^\perp$ in this case is to increase the splitting of the bands around the $K$ ($K'$) points and if $V>\Delta_{SO}$, the Mexican hat is flattened (not shown).

\begin{figure}[t]
\includegraphics[width=.5\textwidth]{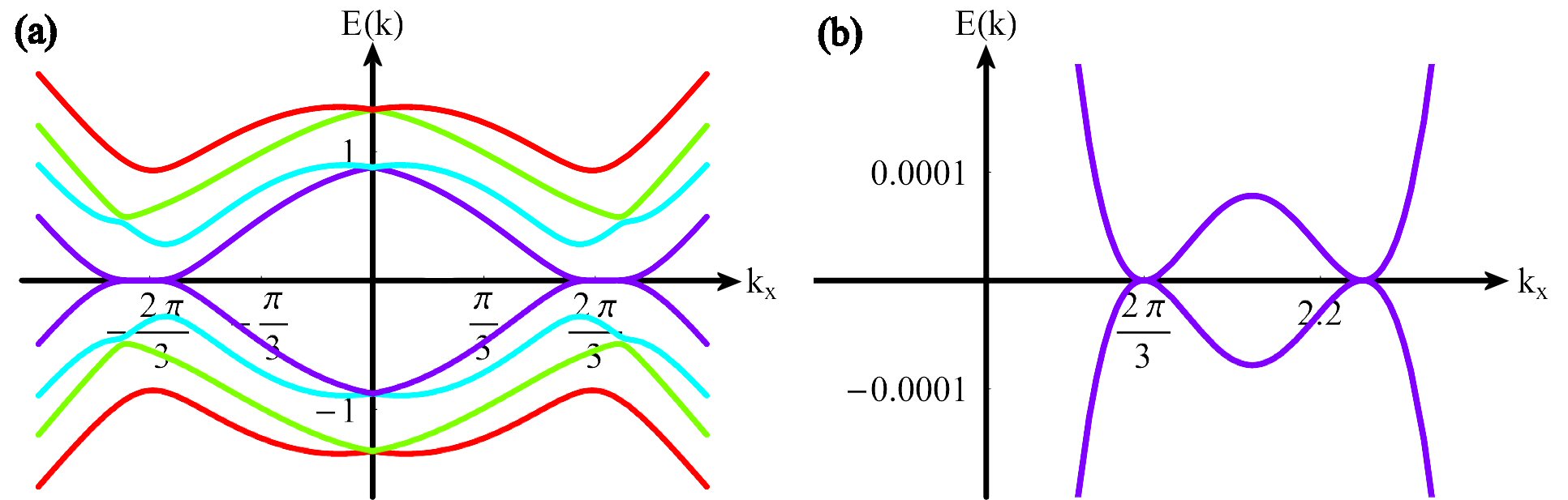}
\caption{(Color online) (a) Energy spectrum of an unbiased system with zero intralayer ISO interaction, but $t_R=0.2$ and $t_R^\perp=0.4$. Compare with Fig.~\ref{Vis0series}(d). (b) Zoom in on the $K$ point of (a). Compare with Fig.~\ref{Vis0serieszi}(a). Other parameters are $t=1$, $t_\perp=0.2$, $a=1$ and $k_y=2 \pi/(\sqrt{3}a)$.}
\label{trinterrash}
\end{figure}

Let us now consider an unbiased system with zero ISO coupling, but with nonzero interlayer and intralayer Rashba coupling. We would have such a system if a tilted electric field is present. The relative strength of both interactions can be tuned independently by changing the parallel and perpendicular components of the electric field. This case is a generalization of Fig.~\ref{Vis0series}(d), which was described in more detail in Figs.~\ref{dsplittingbilayer} and \ref{Vis0serieszi}. The most striking effect of a nonzero $t_R^\perp$ is the lifting of the spin degeneracies at the $K$ ($K'$) points, with as main consequence the destruction of the Dirac cones (see Fig.~\ref{trinterrash}). The way the lowest laying energy band splits is also effected by the interlayer Rashba coupling. Recall that the $K$ ($K'$) point splits into four due to intralayer Rashba coupling and that the central point had a linear crossing at very low energy scales. This linear crossing is modified to a higher order crossing by the interlayer Rashba interaction, as it can be seen in Fig.~\ref{trinterrash}. Note that the energy scale of the zoom in at the $K$ point in Fig.~\ref{trinterrash} has decreased by a factor fifty with respect to Fig.~\ref{Vis0serieszi}. This is also due to the interlayer Rashba coupling.

If all parameters are nonzero the spectrum is very complicated. Interlayer Rashba coupling does deform this spectrum, but we could not detect any special feature that would justify exhibiting them.

\section{Conclusions}
\label{conclusions}

We studied a graphene bilayer including both, the intrinsic and the Rashba SO interactions and we found that these interactions can modify the energy dispersion in a non trivial way.

First, we concentrated on the unbiased system and considered only intralayer interactions. We observed that ISO interactions can open a gap in the system, in the same way as they do for monolayer graphene.\cite{KaMe05b} On the other hand, if only an intralayer Rashba SO interaction is present, the energy bands are completely different than for monolayer graphene. The $K$ ($K'$) points still split into four, but at very low energies the off center points have now a higher order crossing instead of the linear crossing present in monolayer graphene (Fig.~\ref{Vis0serieszi}(a)). The most striking feature, however, is the formation of a Dirac cone out of an energy band that once was parabolic (Fig.~\ref{dsplittingbilayer}(b)). This Dirac cone is located exactly at the former $K$ ($K'$) point. The intralayer Rashba interaction not only lifts the spin degeneracy of the energy bands, but also changes their individual behavior. If both the intralayer ISO and Rashba terms are nonzero, the Dirac cone is destroyed and the particle-hole symmetry is broken.

The presence of a Dirac cone in unbiased bilayer graphene is a very special feature. Together with the split $K$ ($K'$) point, the spectrum that we found will give rise to two different low energy excitations, one of which is massless. The speed with which this massless excitation travels depends on the Rashba constant (see Fig.~\ref{Vis0serieszi}(b)), which can be tuned by the perpendicular electric field. We expect that the dispersion relation shown in Fig.~\ref{Vis0serieszi}(a) could be observed with ARPES experiments, which have already successfully demonstrated the Dirac dispersion in monolayer graphene.\cite{Lan06}

In a biased system with nonzero ISO interactions, the spin degeneracy of the energy bands is lifted. The splitting occurs only around the $K$ ($K'$) points. We also observe a band crossing at the $k_x=0$ point (Fig.~\ref{Vis025series}(c) and (d)). A bias voltage, in combination with the intralayer Rashba coupling, destroys the Dirac cones and the spectrum becomes asymmetric around the $K$ ($K'$) points.

Next, we considered an interlayer Rashba interaction between the planes. This interlayer interaction would in principle be expected to be small compared to the intralayer one because of the larger interlayer atomic separation. However, this effect could still be important if pressure is applied to approach the two layers. In our model, the interlayer Rashba coupling finds its origin in the presence of a tilted electric field. In a bilayer system with no intralayer SO interactions, this interaction causes effectively a shift of the interlayer hopping parameter. However, in a system where the intralayer SO couplings are nonzero, we see a clear effect in the energy spectrum. In an unbiased system with nonzero $\Delta_{SO}$, the spin degeneracy of the bands is lifted around the $K$ ($K'$) points and a Mexican hat feature appears. The Mexican hat feature is known to arise in the bilayer system in the presence of a bias voltage.\cite{CaNe06} Here, however, we found that it can also appear without a bias, but solely due to SO interactions. Moreover, the Mexican hat in the energy dispersion is fully spin polarized. If $\Delta_{SO}$ becomes large enough, the degeneracy of the two lowest laying conduction bands at the $K$ ($K'$) points is shifted to the two middle bands. The system then becomes isospin degenerate. Indeed, in the presence of tunneling, a bilayer can be described as a two-level system, where the energy bands in each layer have combined into symmetric and antisymmetric energy bands, separated by a gap given by the tunneling energy. If we represent the asymmetric band by an isospin up and the symmetric one by an isospin down, we see that the ISO interaction can lead to an isospin degeneracy at the $K$ and $K'$ points, although the spins remain fully polarized.

If the ISO interaction is absent, but the intralayer Rashba is nonzero, we have seen in Fig.~\ref{trinterrash} that the effect of the interlayer Rashba interaction is to destroy the Dirac cone at the $K$ ($K'$) point and to modify the way the $K$ ($K'$) point splits into four. We can no longer observe a linear crossing for the central point and the energy scales associated with this splitting are substantially smaller.

Finally, we would like to discuss the possibility of observing experimentally the effects that we have described above. In monolayer graphene, the current estimates are that the ISO interaction is very small ($0.0011-0.05$ meV).\cite{Tri07} However, it is already possible to tune the Rashba coupling in a graphene layer on a Ni substrate up to $t_R \sim 0.2$ eV.\cite{Yu08} Because these values should be representative for bilayer graphene as well, we expect that the results we found involving the Rashba interaction are well within the experimental reach. We think it should be possible to detect the Dirac cone that arises from the intralayer Rashba term in an unbiased bilayer. If this Dirac cone can be detected and if it is destroyed by an in-plane electric field, we would have an indication that indeed an interlayer Rashba interaction is present in the system.

The values we used for $\Delta_{SO}$ are larger than indicated experimentally. However, we should recall that the same kind of system can be engineered using cold atoms in optical lattices, and in this case there is much less constraint on the parameters of the model. Our aim here was mainly to draw a comparison of the different effects of the Rashba and ISO interactions to determine the trend introduced by each one.

A next step would be to include edge states in the model. If we regard the intralayer ISO interaction and the bias voltage in bilayer graphene as being comparable with the ISO interaction and a staggered sublattice potential in monolayer graphene, there is a possibility that the bilayer system would exhibit a phase transition, equivalent to the one described in Ref~\onlinecite{KaMe05a}. It is already known that there are two types of edge states in bilayer graphene,\cite{CaNe08b} but SO interactions have not yet been taken into account. We hope that our results will motivate further theoretical studies and experiments in the field.

\section*{ACKNOWLEDGMENTS}
We acknowledge financial support from the Netherlands Organization for Scientific Research (NWO). We are grateful to A.H. Castro Neto, N. Sandler, W. Beugeling, and
A. Lazarides for fruitful discussions.

\end{document}